\documentclass[twocolumn,prc,showpacs,showkeys]{revtex4}
\usepackage{amsfonts}
\usepackage{amsmath}
\usepackage{amssymb}
\usepackage{graphicx}
\usepackage{rotating}

\providecommand{\U}[1]{\protect\rule{.1in}{.1in}}
\providecommand{\U}[1]{\protect\rule{.1in}{.1in}}
\providecommand{\U}[1]{\protect\rule{.1in}{.1in}}

\begin{document}

\preprint{}
\title{Forbidden Nuclear Reactions}
\author{P\'{e}ter K\'{a}lm\'{a}n}
\author{Tam\'{a}s Keszthelyi}
\affiliation{Budapest University of Technology and Economics, Institute of Physics,
Budafoki \'{u}t 8. F., H-1521 Budapest, Hungary\ }
\keywords{nuclear reactions: specific reactions: general, quantum mechanics,
fusion reactions}
\pacs{24.90.+d, 03.65.-w, 25.60.Pj}

\begin{abstract}
Exothermal nuclear reactions which become forbidden due to Coulomb repulsion
in the\ $\varepsilon \rightarrow 0$ limit ($\lim_{\varepsilon \rightarrow
0}\sigma \left( \varepsilon \right) =0$) are investigated. ($\sigma \left(
\varepsilon \right) $ is the cross section and $\varepsilon $ is the center
of mass energy.) It is found that \textit{any perturbation} may mix states
with small but finite amplitude to the initial state resulting finite cross
section (and rate) of the originally forbidden nuclear reaction in the $%
\varepsilon \rightarrow 0$ limit. The statement is illustrated by
modification of nuclear reactions due to impurities in a gas mix of atomic
state. The change of the wavefunction of reacting particles in nuclear range
due to their Coulomb interaction with impurity is determined using standard
time independent perturbation calculation of quantum mechanics. As an
example, cross section, rate and power densities of impurity assisted
nuclear $pd$ reaction are numerically calculated. With the aid of
astrophysical factors cross section and power densities of the impurity
assisted $d(d,n)_{2}^{3}He$, $d(d,p)t$, $d(t,n)_{2}^{4}He$, $%
_{2}^{3}He(d,p)_{2}^{4}He$, $_{3}^{6}Li(p,\alpha )_{2}^{3}He$, $%
_{3}^{6}Li(d,\alpha )_{2}^{4}He$, $_{3}^{7}Li(p,\alpha )_{2}^{4}He$, $%
_{4}^{9}Be(p,\alpha )_{3}^{6}Li$, $_{4}^{9}Be(p,d)_{4}^{8}Be$, $%
_{4}^{9}Be(\alpha ,n)_{6}^{12}C$, $_{5}^{10}B(p,\alpha )_{4}^{7}Be$ and $%
_{5}^{11}B(p,\alpha )_{4}^{8}Be$ reactions are also given. The affect of gas
mix-wall interaction on the process is considered too.
\end{abstract}

\volumenumber{number}
\issuenumber{number}
\eid{identifier}
\date[Date text]{date}
\received[Received text]{date}
\revised[Revised text]{date}
\accepted[Accepted text]{date}
\published[Published text]{date}
\startpage{1}
\endpage{}
\maketitle

\section{Introduction}

The cross section $\left( \sigma \right) $ of nuclear reactions between
charged particles $j$ and $k$ of charge numbers $z_{j}$ and $z_{k}$ reads as
\cite{Angulo}
\begin{equation}
\sigma \left( \varepsilon \right) =S\left( \varepsilon \right) \exp \left[
-2\pi \eta _{jk}\left( \varepsilon \right) \right] /\varepsilon ,
\label{sigma}
\end{equation}%
where $S\left( \varepsilon \right) $ is the astrophysical $S$-factor and $%
\varepsilon $ is the kinetic energy taken in the center of mass $\left(
CM\right) $ coordinate system.%
\begin{equation}
\eta _{jk}=z_{j}z_{k}\alpha _{f}\frac{a_{jk}m_{0}c}{\hbar \left\vert \mathbf{%
k}\right\vert }=z_{j}z_{k}\alpha _{f}\sqrt{a_{jk}\frac{m_{0}c^{2}}{%
2\varepsilon }}  \label{etajk}
\end{equation}%
is the Sommerfeld parameter, where $\mathbf{k}$ is the wave number vector of
particles $j$ and $k$ in their relative motion, $\hbar $ is the reduced
Planck-constant, $c$ is the velocity of light in vacuum and%
\begin{equation}
a_{jk}=\frac{A_{j}A_{k}}{A_{j}+A_{k}}  \label{ajk}
\end{equation}%
is the reduced mass number of particles $j$ and $k$ of mass numbers $A_{j}$
and $A_{k}$ and rest masses $m_{j}=A_{j}m_{0}$, $m_{k}=A_{k}m_{0}$. $%
m_{0}c^{2}=931.494$ MeV is the atomic energy unit, $\alpha _{f}$ is the fine
structure constant.

However in the latest few decades "anomalies" to (\ref{sigma}) were
reported, which are anomalous screening effect and the less well documented
area of phenomena of the so called low energy nuclear reactions (LENR).

Extraordinary observations in cross section measurements of $dd$ reactions
in deuterated metal targets made in low energy accelerator physics which can
not be explained by electron screening are named anomalous screening.
(Systematic survey of anomalous screening effect was made \cite{Huke} a
decade ago.) However the full theoretical explanation of the effect is still
missing.

In low-energy nuclear reactions (LENR), a new and problematic field that
emerged after the notorious "cold fusion" publication by Fleischmann and
Pons in 1989 \cite{FP1}, results are reported that are in conflict with (\ref%
{sigma}). Despite the fact that even the possibility of the phenomenon of
nuclear fusion at low energies is received with due scepticism in mainstream
physics \cite{Huizenga} low-energy nuclear reactions (LENR) are dealt with
in a great number of laboratories and publications (mostly experimental),
conferences and periodicals have been devoted to various aspects of the
problem. (For summary of the field see e.g. \cite{Krivit}, \cite{Storms3},
\cite{Storms2}, \cite{Storms1}.)

The aim of this paper is to show the possible reason for anomalies of cross
sections of nuclear reactions of particles of like charges at low energy in
general.

\section{Role of Coulomb repulsion}

The solution $\varphi _{jk}\left( \mathbf{R,r}\right) $ of the sationary Schr%
\"{o}dinger equation
\begin{equation}
H_{jk}\varphi _{jk}\left( \mathbf{R,r}\right) =E_{jk}\varphi _{jk}\left(
\mathbf{R,r}\right)  \label{Hjk}
\end{equation}%
of particles of charge numbers $z_{j}$ and $z_{k}$ with
\begin{equation}
H_{jk}=-\frac{\hbar ^{2}}{2m_{0}\left( A_{j}+A_{k}\right) }\nabla _{\mathbf{R%
}}^{2}-\frac{\hbar ^{2}}{2m_{0}a_{jk}}\nabla _{\mathbf{r}}^{2}+\frac{%
z_{j}z_{k}e^{2}}{\left\vert \mathbf{r}\right\vert }  \label{Hjk2}
\end{equation}%
is
\begin{equation}
\varphi _{jk}\left( \mathbf{R,r}\right) =V^{-1/2}e^{i\mathbf{KR}}\varphi _{%
\text{Cb}}(\mathbf{r}),  \label{fijk}
\end{equation}%
where $\mathbf{R}=\left( m_{j}\mathbf{r}_{j}+m_{k}\mathbf{r}_{k}\right)
/\left( m_{j}+m_{k}\right) $ and $\mathbf{r=r}_{jk}=\mathbf{r}_{j}-\mathbf{r}%
_{k}$ are $CM$ and relative coordinate of particles $j$ and $k$ of
coordinate $\mathbf{r}_{j}$ and $\mathbf{r}_{k}$, respectively. $V$ denotes
the volume of normalization and $\varphi _{\text{Cb}}(\mathbf{r})$ is the
Coulomb solution \cite{Alder}, which is the wavefunction of the relative
motion in repulsive Coulomb potential. $\nabla _{\mathbf{R}}^{2}$ and $%
\nabla _{\mathbf{r}}^{2}$ are Laplace operators in the $CM$ and relative
coordinates, $\mathbf{K}$ is the wave vector of the $CM$ motion and $%
E_{jk}=E_{\text{CM}}+\varepsilon $ with $E_{\text{CM}}=\hbar ^{2}\mathbf{K}%
^{2}/\left[ 2m_{0}\left( A_{j}+A_{k}\right) \right] $ and $\varepsilon
=\hbar ^{2}\mathbf{k}^{2}/\left( 2m_{0}a_{jk}\right) $. $e$ is the
elementary charge with $e^{2}=\alpha _{f}\hbar c$.

The contact probability density in the nuclear volume is $\left\vert \varphi
_{Cb}(\mathbf{0})\right\vert ^{2}=f_{jk}^{2}/V$, where
\begin{equation}
f_{jk}=\left\vert e^{-\pi \eta _{jk}/2}\Gamma (1+i\eta _{jk})\right\vert =%
\sqrt{\frac{2\pi \eta _{jk}}{\exp \left( 2\pi \eta _{jk}\right) -1}}.
\label{Fjk}
\end{equation}%
The result of a first order calculation of the cross section in standard
perturbation theory of quantum mechanics is proportional to $f_{jk}^{2}/v$
where $v$ is the relative velocity in the $CM$ system. Investigating the
energy dependence of $f_{jk}^{2}/v$ it is found that $f_{jk}^{2}/v\sim \exp %
\left[ -2\pi \eta _{jk}\left( \varepsilon \right) \right] /\varepsilon $ in
the $\varepsilon \rightarrow 0$ limit. Accordingly, the magnitude of the
factor $f_{jk}$ is crucial from the point of view of magnitude of the cross
section.

If the reaction energy $\Delta >0$ (the difference between initial and final
rest energies) of reaction between particles of likewise charge, the
spontaneous process could be allowed by energy conservation. However in the $%
\varepsilon \rightarrow 0$ limit $\lim_{\varepsilon \rightarrow
0}f_{jk}^{2}\left( \varepsilon \right) =0$ with $\lim_{\varepsilon
\rightarrow 0}\left\vert \varphi _{\text{Cb}}(\mathbf{0})\right\vert ^{2}=0$
and the process \textit{becomes forbidden} ($\lim_{\varepsilon \rightarrow
0}\sigma \left( \varepsilon \right) =0$) due to Coulomb repulsion. (If one
of the reacting particles is neutral, which is the case of neutron capture
processes, the cross section has non zero value in the $\varepsilon
\rightarrow 0$ limit, see e.g. thermal neutron absorption cross sections
\cite{Blatt}.)

\section{Statement and examples}

Experience in atomic physics indicates that in case of forbidden transitions
the second order process may play an important role. As e.g. in the case of
the hydrogen $2s_{1/2}-1s_{1/2}$ transition, which is a forbidden electric
dipole transition, the largest transition rate comes from a two photonic
process \cite{Bethe} in which the sum of the energies of the simultaneously
emitted photons equals the difference between the energies of states $%
2s_{1/2}$ and $1s_{1/2}$. The mean life time $1/7$ s of the $2s_{1/2}$ state
due to the two photonic process is much longer than the lifetime $1.6\times
10^{-9}$ s of state $2p_{1/2}$ for which electric dipole transition is
allowed. Thus one can conclude that a second order process from the point of
view of perturbation calculation can result small but finite transition
rate. In the second order process the state is changed in first order and
states, which can produce allowed electric dipole transition rate, are mixed
with small amplitude to the initial $2s_{1/2}$ state meanwhile two particles
are emitted.

Similarly an essential change of the initial eigenstate of $\left( \ref{Hjk}%
\right) $ of $\varepsilon =0$ may happen due to \textit{any perturbation}
since it can mix states of $\varepsilon \neq 0$ with small but finite
amplitude to the initial state resulting much smaller (compared to neutron
absorption) but finite rate of the nuclear reaction originally forbidden in
the $\varepsilon \rightarrow 0$ limit. Consequently, cross section and rate
of processes to be considered should be calculated by the rules of standard
perturbation calculation of quantum mechanics. Our statement applies to
every nuclear process for which $\sigma \left( \varepsilon \right) $ has the
form of $\left( \ref{sigma}\right) $ and $\lim_{\varepsilon \rightarrow
0}\sigma \left( \varepsilon \right) =0$ holds, and as such it concerns low
energy nuclear physics with charged participants in general.

Since the above statement is quite general it is only illustrated by
modification of forbidden nuclear reactions due to Coulomb interaction with
impurities (the initial state is defined in the next section). We
demonstrate the mechanism on the%
\begin{equation}
_{z_{1}}^{A_{1}}V+\text{ }_{z_{2}}^{A_{2}}w+\text{ }_{z_{3}}^{A_{3}}X%
\rightarrow \text{ }_{z_{1}}^{A_{1}}V^{\prime }+\text{ }%
_{z_{3}+z_{2}}^{A_{3}+A_{2}}Y+\Delta  \label{Reaction 3}
\end{equation}%
and
\begin{equation}
_{z_{1}}^{A_{1}}V+\text{ }_{z_{2}}^{A_{2}}w+\text{ }_{z_{3}}^{A_{3}}X%
\rightarrow \text{ }_{z_{1}}^{A_{1}}V^{\prime }+\text{ }_{z_{4}}^{A_{4}}Y+%
\text{ }_{z_{5}}^{A_{5}}W+\Delta  \label{Reaction 4}
\end{equation}%
processes. Reaction $\left( \ref{Reaction 3}\right) $ is an impurity $\left(
_{z_{1}}^{A_{1}}V\right) $ assisted capture of particle $_{z_{2}}^{A_{2}}w$,
e.g. capture of proton $\left( p\right) $, deuteron $\left( d\right) $,
triton $\left( t\right) $, $^{3}He$, $^{4}He$, etc. The impurity $\left(
_{z_{1}}^{A_{1}}V\right) $ assisted reaction $\left( \ref{Reaction 4}\right)
$ with two final fragments is possible with conditions $%
A_{2}+A_{3}=A_{4}+A_{5}$ and $z_{2}+z_{3}=z_{4}+z_{5}$. The reaction energy $%
\Delta $ is the difference between the sum of the initial and final mass
excesses, i.e. $\Delta =\Delta _{A_{2},z_{2}}+\Delta _{A_{3},z_{3}}-\Delta
_{A_{3}+A_{2},z_{3}+z_{2}}$ in case of $\left( \ref{Reaction 3}\right) $ and
$\Delta =\Delta _{A_{2},z_{2}}+\Delta _{A_{3},z_{3}}-\Delta
_{A_{4},z_{4}}-\Delta _{A_{5},z_{5}}$ in case of $\left( \ref{Reaction 4}%
\right) $ where $\Delta _{A_{j},z_{j}}$ and $\Delta
_{A_{3}+A_{2},z_{3}+z_{2}}$ are the corresponding mass excesses \cite{Shir}.
Since particle $1$ merely assists the nuclear reaction its rest mass does
not change.

Usually capture of particle $_{z_{2}}^{A_{2}}w$ may happen in the $%
_{z_{2}}^{A_{2}}w+$ $_{z_{3}}^{A_{3}}X\rightarrow $ $%
_{z_{3}+z_{2}}^{A_{3}+A_{2}}Y+\gamma $ (with $\Delta >0$) reaction where $%
\gamma $ emission is required by energy and momentum conservation.
Accordingly $\left( \ref{Reaction 3}\right) $ describes a new type of $%
_{z_{2}}^{A_{2}}w$-capture. In the usual $_{z_{2}}^{A_{2}}w$-capture
reaction\ particles $_{z_{3}+z_{2}}^{A_{3}+A_{2}}Y$ and $\gamma $ take away
the reaction energy and the reaction is governed by electromagnetic
interaction. In reaction $\left( \ref{Reaction 3}\right) $ the reaction
energy is taken away by particles $_{z_{1}}^{A_{1}}V^{\prime }$ and $%
_{z_{3}+z_{2}}^{A_{3}+A_{2}}Y$ while the reaction is governed by Coulomb as
well as strong interactions.

\section{Mechanism and model}

It is assumed that initially all components of a 3-body system are in atomic
state. Atomic state can effectively be achieved e.g. by dissociative
chemisorption at metal (e.g. $Pd$, $Ni$ and $Cu$) surfaces from two atomic
molecules \cite{Kroes} or simply by heating a molecular gas. So, as initial
system three screened charged heavy particles of rest masses $m_{j}$ and
nuclear charges $z_{j}e$ ($j=1,2,3$) are taken. The total Hamiltonian which
describes this 3-body system is
\begin{equation}
H_{\text{tot}}=H_{\text{kin,1}}+H_{\text{23,sc}}+V_{\text{Cb,sc}}(1,2)+V_{%
\text{Cb,sc}}(1,3),  \label{Htot}
\end{equation}%
where $H_{\text{23,sc}}=H_{\text{kin,2}}+H_{\text{kin,3}}+V_{\text{Cb,sc}%
}(2,3)$ is the Hamiltonian of particles $2$ and $3$ whose nuclear reaction
will be discussed. $H_{\text{kin,j}}$ denotes the kinetic Hamiltonian of
particle $j$ and particle $1$ is considered to be free.
\begin{equation}
V_{\text{Cb,sc}}\left( j,k\right) =\frac{z_{j}z_{k}e^{2}}{2\pi ^{2}}\int
\frac{\exp (i\mathbf{qr}_{jk}\mathbf{)}}{q^{2}+q_{\text{sc,jk}}^{2}}d\mathbf{%
q,}  \label{Vcb1}
\end{equation}%
denotes the screened Coulomb interaction between particles $j$ and $k$ with
screening parameter $q_{\text{sc,jk}}$.

It is supposed that stationary solutions $\left\vert 1\right\rangle $ and $%
\left\vert 2,3\right\rangle _{\text{sc}}$ of energy eigenvalues $E_{1}\emph{%
\ }$and $E_{23}$ of the stationary Schr\"{o}dinger equations $H_{\text{kin,1}%
}\left\vert 1\right\rangle =E_{1}\left\vert 1\right\rangle $ with $E_{1}$
the kinetic energy of particle $1$ and $H_{\text{23,sc}}\left\vert
2,3\right\rangle _{\text{sc}}=E_{23}\left\vert 2,3\right\rangle _{\text{sc}}$
with $E_{23}=E_{\text{CM}}+\varepsilon $ are known. Here $\varepsilon $ and $%
E_{\text{CM}}$ are the energies attached to the relative and $CM$ motions
(of wave numbers $\mathbf{k}$ and $\mathbf{K}$) of particles $2$ and $3$.
Thus $H_{\text{tot}}$ can be written as $H_{\text{tot}}=H_{0}+H_{\text{Int}}$
with $H_{0}=H_{1}+H_{\text{23,sc}}$ as the unperturbed Hamiltonian and%
\begin{equation}
H_{\text{Int}}=V_{\text{Cb,sc}}(1,2)+V_{\text{Cb,sc}}(1,3)  \label{Hint}
\end{equation}%
as the interaction Hamiltonian (time independent perturbation). The
stationary solution $\left\vert 1,2,3\right\rangle _{\text{0,sc}}$ of $%
H_{0}\left\vert 1,2,3\right\rangle _{\text{0,sc}}=E_{0}\left\vert
1,2,3\right\rangle _{\text{0,sc}}$ with $E_{0}=E_{1}+E_{23}$ can be written
as $\left\vert 1,2,3\right\rangle _{\text{0,sc}}=\left\vert 1\right\rangle
\left\vert 2,3\right\rangle _{\text{sc}}$ which is the direct product of
states $\left\vert 1\right\rangle $ and $\left\vert 2,3\right\rangle _{\text{%
sc}}$. The states $\left\vert 1,2,3\right\rangle _{\text{0,sc}}$ form
complete system. The approximate solution of $H_{\text{tot}}\left\vert
1,2,3\right\rangle _{\text{sc}}=E_{0}\left\vert 1,2,3\right\rangle _{\text{sc%
}}$ in the screened case is obtained with the aid of standard time
independent perturbation calculation \cite{Landau} and the first order
approximation is expanded in terms, which are called intermediate states, of
the complete system $\left\vert 1,2,3\right\rangle _{\text{0,sc}}$.

The solutions $\left\vert 2,3\right\rangle _{\text{sc}}$ in the screened
case are unknown (their coordinate representation $\langle \mathbf{R,r}%
\left\vert 2,3\right\rangle _{\text{sc}}$ is denoted by $\varphi _{23}\left(
\mathbf{R,r}\right) _{\text{sc}}$) but the solution of $H_{23}\left\vert
2,3\right\rangle =E_{23}\left\vert 2,3\right\rangle $ in the unscreened case
is known and the coordinate representation $\langle \mathbf{R,r}\left\vert
2,3\right\rangle =\varphi _{23}\left( \mathbf{R,r}\right) $ of $\left\vert
2,3\right\rangle $, as it is said above, has the form $\varphi _{23}\left(
\mathbf{R,r}\right) =V^{-1/2}e^{i\mathbf{KR}}\varphi _{\text{Cb}}(\mathbf{r}%
) $, where $\varphi _{\text{Cb}}(\mathbf{r})$ is the unscreened Coulomb
solution \cite{Alder} (now $\mathbf{r}=\mathbf{r}_{23}$).

The two important limits of $\varphi _{23}\left( \mathbf{R,r}\right) _{\text{%
sc}}$\ are: the solution $\varphi _{23}\left( \mathbf{R,r,}nucl\right) _{%
\text{sc}}$ in the nuclear volume and the solution $\varphi _{23}\left(
\mathbf{R,r,}out\right) _{\text{sc}}$ in the screened regime. In the nuclear
volume screening is negligible thus $\varphi _{23}\left( \mathbf{R,r}\right)
_{\text{sc}}=\varphi _{23}\left( \mathbf{R,r}\right) $. Furthermore, in this
case in $\varphi _{23}\left( \mathbf{R,r}\right) $ an approximate form $%
\varphi _{\text{Cb,a}}(\mathbf{r})=e^{i\mathbf{k}\cdot \mathbf{r}%
}f_{23}(\left\vert \mathbf{k}\right\vert )/\sqrt{V}$ of the (unscreened)
Coulomb solution $\varphi _{\text{Cb}}(\mathbf{r})$ may be used. Here $%
f_{23}(\left\vert \mathbf{k}\right\vert )$ is the appropriate factor given
by $\left( \ref{Fjk}\right) $ corresponding to particles $2$ and $3$. Thus $%
\varphi _{23}\left( \mathbf{R,r,}nucl\right) _{\text{sc}}=f_{23}(\left\vert
\mathbf{k}\right\vert )e^{i\mathbf{KR}}e^{i\mathbf{kr}}/V$ is used in the
range of the nucleus and in the calculation of the nuclear matrix-element.
In the screened (outer) range, where Coulomb potential is negligible, the
solution becomes $\varphi _{23}\left( \mathbf{R,r,}out\right) _{\text{sc}%
}=e^{i\mathbf{KR}}e^{i\mathbf{kr}}/V$ that is used in the calculation of the
Coulomb matrix element.

In the screened range the initial wave function of zero energy is $\varphi _{%
\text{i}}=V^{-3/2}$. The intermediate states of particles $2$ and $3$ are
determined by the wave number vectors $\mathbf{K}$ and $\mathbf{k}$. In the
case of the assisting particle $1$ the intermediate and final state is a
plane wave of wave number vector $\mathbf{k}_{1}$.

The matrix elements $V_{\text{Cb,}\nu \text{i}}$ of the screened Coulomb
potential between the initial and intermediate states are%
\begin{eqnarray}
V_{Cb}(1,s)_{\nu \text{i}} &=&\frac{z_{1}z_{s}}{2\pi ^{2}}e^{2}\frac{\left(
2\pi \right) ^{9}}{V^{3}}\delta \left( \mathbf{k}_{1}+\mathbf{K}\right)
\times  \label{VCb1snui} \\
&&\times \frac{\delta \left( \mathbf{k}+a(s)\mathbf{k}_{1}\right) }{\mathbf{k%
}_{1}^{2}+q_{\text{sc,1s}}^{2}}\text{ \ \ \ \ }  \notag
\end{eqnarray}%
where $a(s)=\left( -A_{3}\delta _{s,2}+A_{2}\delta _{s,3}\right) /\left(
A_{2}+A_{3}\right) $ \ and $s=2,3$.

\section{Change of three-particle wavefunction\textit{\ }in nuclear range}

According to standard time independent perturbation theory of quantum
mechanics \cite{Landau} the first order change of the wavefunction in the
range $r\lesssim R_{0}$ ($R_{0}$ is the nuclear radius of particle $3$) due
to screened Coulomb perturbation is determined as%
\begin{equation}
\delta \varphi \left( \mathbf{r}\right) =\sum_{s=2,3}\delta \varphi \left( s,%
\mathbf{r}\right)  \label{dfi}
\end{equation}%
with%
\begin{eqnarray}
\delta \varphi \left( s,\mathbf{r}\right) &=&\int \int \frac{V_{\text{Cb}%
}(1,s)_{\nu \text{i}}}{E_{\nu }-E_{i}}\frac{V}{\left( 2\pi \right) ^{6}}%
\times  \label{dfis} \\
&&\times e^{i(\mathbf{KR}+\mathbf{\mathbf{k}_{1}\mathbf{r}_{1})}}\varphi _{%
\text{Cb,a}}(\mathbf{k},\mathbf{r})d\mathbf{K}d\mathbf{k},  \notag
\end{eqnarray}%
where $E_{i}$ and $E_{\nu }$ are the kinetic energies in the initial and
intermediate states, respectively. The initial momenta and kinetic energies
of particles $1$, $2$ and $3$ are zero $\left( E_{i}=0\right) $ and $E_{\nu
}=E_{23}+\hbar ^{2}\mathbf{k}_{1}^{2}/\left( 2m_{0}A_{1}\right) $. Thus%
\begin{eqnarray}
\delta \varphi \left( s,\mathbf{r}\right) &=&z_{1}z_{s}\alpha _{f}\frac{4\pi
\hbar c}{V^{5/2}}\frac{e^{i(\mathbf{\mathbf{k}_{1}\mathbf{r}_{1}-k}_{1}%
\mathbf{R)}}}{\mathbf{k}_{1}^{2}+q_{\text{sc,1s}}^{2}}\times  \label{dfi2} \\
&&\times \frac{2m_{0}a_{1s}}{\hbar ^{2}\mathbf{k}_{1}^{2}}\left[
f_{23}\left( k\right) e^{i\mathbf{kr}}\right] _{\mathbf{k}=a(s)\mathbf{k}%
_{1}}.  \notag
\end{eqnarray}%
It can be seen that the arguments of $f_{23}\left( \left\vert \mathbf{k}%
\right\vert \right) $ are $\left\vert \mathbf{k}\right\vert =\frac{A_{3}}{%
A_{2}+A_{3}}k_{1}$ and $\left\vert \mathbf{k}\right\vert =\frac{A_{2}}{%
A_{2}+A_{3}}k_{1}$, here $k_{1}=\left\vert \mathbf{k}_{1}\right\vert $.
Consequently, if particle $1$ obtains large kinetic energy, as is the case
in nuclear reactions (e.g. $\mathbf{k}_{1}^{2}=k_{0}^{2}=2m_{0}a_{14}\Delta
\hbar ^{-2}$ in the case of reaction $\left( \ref{Reaction 3}\right) $),
then the factors $f_{23}\left( \left\vert \mathbf{k}\right\vert \right) $
and the rate of the process too will be considerable. (In this case one can
neglect $q_{\text{sc,jk}}^{2}$ in the denominator of $\left( \ref{dfi2}%
\right) $). Since $\lim_{\varepsilon \rightarrow 0}\left\vert \delta \varphi
\left( \mathbf{0}\right) \right\vert ^{2}\neq 0$, i.e. it remains finite in
the $\varepsilon \rightarrow 0$ limit, and the expected reaction rate too
remains finite. Furthermore, $\delta \varphi \left( \mathbf{r}\right) $,
which causes the effect, is temperature independent. (Temperature dependence
is brought in by mechanisms responsible for producing atomic states.) Up to
this point the calculation and the results are nuclear reaction and nuclear
model independent.

\section{Cross section}

When calculating the cross section of reaction $_{z_{1}}^{A_{1}}V+p+d%
\rightarrow $ $_{z_{1}}^{A_{1}}V^{\prime }+$ $_{2}^{3}He+5.493$ MeV the
Hamiltonian $V_{\text{st}}\left( 2,3\right) =-V_{0}$ \ if $\left\vert
\mathbf{r}_{23}\right\vert =\left\vert \mathbf{r}\right\vert \leq b$ and $V_{%
\text{st}}\left( 2,3\right) =0$ \ if $\left\vert \mathbf{r}_{23}\right\vert
=\left\vert \mathbf{r}\right\vert >b$ of strong interaction which is
responsible for nuclear reaction between particles $2$ and $3$ is used. For
the final state of the captured proton the Weisskopf-approximation is
applied, i.e. $\Phi _{f}\left( \mathbf{r}\right) =\Phi _{\text{fW}}\left(
\mathbf{r}\right) $ with $\Phi _{\text{fW}}\left( \mathbf{r}\right) =\sqrt{%
3/\left( 4\pi R_{0}^{3}\right) }$ if $r\leq R_{0}$, and $\Phi _{\text{fW}%
}\left( \mathbf{r}\right) =0$ for $r>R_{0}$, where $R_{0}$ is the nuclear
radius. We take $V_{0}=25$ MeV and $R_{0}=b=2\times 10^{-13}$ cm \cite{Blatt}
in the case of $pd$ reaction.

The matrix element of the potential of the strong interaction between
intermediate $\left( e^{i\mathbf{KR}}\varphi _{\text{Cb,a}}(\mathbf{k},%
\mathbf{r})/\sqrt{V}\right) $ and final $\left( e^{i\mathbf{k}_{4}\cdot
\mathbf{R}}\Phi _{\text{f}}\left( \mathbf{r}\right) /\sqrt{V}\right) $
states and in the Weisskopf-approximation is
\begin{equation}
V_{\text{st,f}\nu }^{\text{W}}=-V_{0}\frac{\sqrt{12\pi R_{0}}}{k}%
f_{23}(k)H\left( k\right) \frac{\left( 2\pi \right) ^{3}}{V^{3/2}}\delta
\left( \mathbf{K}-\mathbf{k}_{4}\right)
\end{equation}%
where $H\left( k\right) =\int_{0}^{1}\sin (kR_{0}x)xdx$. According to
standard time independent perturbation theory of quantum mechanics \cite%
{Landau} the transition probability per unit time $\left( W_{\text{fi}}^{%
\text{(2)}}\right) $ of the process can be written as%
\begin{equation}
W_{\text{fi}}^{\text{(2)}}=\frac{2\pi }{\hbar }\int \int \left\vert T_{\text{%
fi}}^{\text{(2)}}\right\vert ^{2}\delta (E_{f}-\Delta )\frac{V^{2}}{\left(
2\pi \right) ^{6}}d\mathbf{k}_{1}d\mathbf{k}_{4}  \label{Wfie}
\end{equation}%
with%
\begin{equation}
T_{\text{fi}}^{\text{(2)}}=\int \int \sum_{s=2,3}\frac{V_{\text{st,f}\nu }V_{%
\text{Cb}}(1,s)_{\nu \text{i}}}{E_{\nu }-E_{i}}\frac{V^{2}}{\left( 2\pi
\right) ^{6}}d\mathbf{K}d\mathbf{k}.  \label{Tif}
\end{equation}

Substituting everything obtained above into $\left( \ref{Tif}\right) $ and $%
\left( \ref{Wfie}\right) $, where $E_{f}$ is the sum of kinetic energies of
the final particles ($1$ and $4$), one can calculate $W_{\text{fi}}^{\text{%
(2)}}$. The cross section $\sigma _{23}^{\left( 2\right) }$ of the process
is defined as $\sigma _{23}^{\left( 2\right) }=N_{1}W_{\text{fi}}^{\text{(2)}%
}/\left( v_{23}/V\right) $ where $N_{1}$ is the number of particles $1$ in
the normalization volume $V$ and $v_{23}/V$ is the flux of particle $2$ of
relative velocity $v_{23}$.%
\begin{equation}
v_{23}\sigma _{23}^{\left( 2\right) }=n_{1}S_{\text{pd}}  \label{sigma23-2}
\end{equation}%
where $n_{1}=N_{1}/V$ is the number density of particles $1$ and%
\begin{eqnarray}
S_{\text{pd}} &=&24\pi ^{2}\sqrt{2}cR_{0}\frac{z_{1}^{2}\alpha
_{f}^{2}V_{0}^{2}\left( \hbar c\right) ^{4}}{\Delta ^{9/2}\left(
m_{0}c^{2}\right) ^{3/2}}\times  \label{SRANR} \\
&&\times \frac{\left( A_{2}+A_{3}\right) ^{2}}{a_{14}^{7/2}}\left[ F\left(
2\right) +F\left( 3\right) \right] ^{2}  \notag
\end{eqnarray}%
with%
\begin{equation}
F(s)=\frac{z_{s}a_{1s}}{A_{3}\delta _{s,2}+A_{2}\delta _{s,3}}f_{23}\left[
a(s)k_{0}\right] H\left[ a(s)k_{0}\right] ,  \label{Fs}
\end{equation}%
$s=2,3$ and $k_{0}=\hbar ^{-1}\sqrt{2m_{0}a_{14}\Delta }$.

In the case of reactions with two final fragments (see $\left( \ref{Reaction
4}\right) $) the nuclear matrix element can be derived from $S(\varepsilon )$
(see $\left( \ref{sigma}\right) $), i.e. in long wavelength approximation
from $S(0)$ which is the astrophysical $S$-factor at $\varepsilon =0$, in
the following manner.

Calculating the transition probability per unit time $W_{\text{fi}}^{\text{%
(1)}}$ of the usual (first order) process in standard manner
\begin{equation}
W_{\text{fi}}^{\text{(1)}}=\int \frac{2\pi }{\hbar }\left\vert V_{\text{st,fi%
}}\right\vert ^{2}\delta \left( E_{f}-\Delta \right) \frac{V}{\left( 2\pi
\right) ^{3}}d\mathbf{k}_{f},
\end{equation}%
where $\mathbf{k}_{f}$ is the relative wave number of the two fragments of
rest masses $m_{4}=m_{0}A_{4}$, $m_{5}=m_{0}A_{5}$ and atomic numbers $A_{4}$%
, $A_{5}$, and $E_{f}=$ $\hbar ^{2}\mathbf{k}_{f}^{2}/(2m_{0}a_{45})$ is the
sum of their kinetic energy. For the magnitude of nuclear matrix element $V_{%
\text{st,fi}}$ we take the form $\left\vert V_{\text{st,fi}}\right\vert
=f_{23}\left( k_{i}\right) \left\vert h_{\text{fi}}\right\vert /V$, where $%
f_{23}\left( k_{i}\right) $ is the Coulomb factor of the initial particles $%
2 $ and $3$ with $k_{i}$ the magnitude of their relative wave number vector $%
\mathbf{k}_{i}$. (The Coulomb factor $f_{45}\left( k_{f}\right) \approx 1$
of the final particles $4$ and $5$ with $k_{f}$ the magnitude of their
relative wave number vector $\mathbf{k}_{f}$.) It is supposed that $%
\left\vert h_{\text{fi}}\right\vert $ does not depend on $\mathbf{k}_{i}$
and $\mathbf{k}_{f}$ namely the long wavelength approximation is used. In
this case the product of the relative velocity $v_{23}$ of the initial
particles $2$, $3$\emph{\ }and the cross section $\sigma _{23}^{(1)}$ is
\begin{equation}
v_{23}\sigma _{23}^{\left( 1\right) }=\frac{\left\vert h_{\text{fi}%
}\right\vert ^{2}f_{23}^{2}\left( k_{i}\right) \left( m_{0}a_{45}\right)
^{3/2}\sqrt{2\Delta }}{\pi \hbar ^{4}}.
\end{equation}%
On the other hand, $v_{23}\sigma _{23}^{\left( 1\right) }$ is expressed with
the aid of $\left( \ref{sigma}\right) $ and $v_{23}=\sqrt{2\varepsilon
/\left( m_{0}a_{23}\right) }$. From the equality of the two kinds of $%
v_{23}\sigma _{23}^{\left( 1\right) }$ one gets
\begin{equation}
\left\vert h_{\text{fi}}\right\vert ^{2}=\frac{\left( \hbar c\right) ^{4}S(0)%
}{z_{2}z_{3}\alpha _{f}\left( m_{0}c^{2}\right) ^{5/2}\sqrt{2\Delta }%
a_{45}^{3/2}a_{23}}.
\end{equation}

In the case of the impurity assisted, second order process $\left\vert V_{%
\text{st,f}\nu }\right\vert =f_{23}\left( k\right) \left\vert h_{\text{fi}%
}\right\vert \left( 2\pi \right) ^{3}\delta \left( \mathbf{K}-\mathbf{K}%
_{f}\right) /V^{2}$ where $\mathbf{K}_{f}$ and $\mathbf{k}_{f}$ are the
final wave number vectors attached to $CM$ and relative motions of the two
final fragments, particles $4$ and $5$. $\mathbf{k}_{f}$ appears in $E_{f}$\
in the energy Dirac-delta. Repeating the calculation of the transition
probability per unit time of the impurity assisted, second order process
applying the above expression of $\left\vert V_{\text{st,f}\nu }\right\vert $
one gets%
\begin{equation}
v_{23}\sigma _{23}^{\left( 2\right) }=n_{1}S_{\text{reaction}},
\label{v23sigma23-2}
\end{equation}%
where $\sigma _{23}^{\left( 2\right) }$ is the cross section of the process
and%
\begin{equation}
S_{\text{reaction}}=\frac{8\alpha _{f}^{2}z_{1}^{2}}{a_{23}a_{123}^{3}}\frac{%
S(0)c}{m_{0}c^{2}}\left( \frac{\hbar c}{\Delta }\right) ^{3}I
\label{result2}
\end{equation}%
with%
\begin{equation}
I=\int_{0}^{1}\left( \sum_{s=2,3}\frac{z_{s}a_{1s}\sqrt{A_{s}}}{\sqrt{%
e^{b_{23}A_{s}\frac{1}{x}}-1}}\right) ^{2}\frac{\sqrt{1-x^{2}}}{x^{7}}dx.
\label{Iintcharged}
\end{equation}%
Here $b_{23}=2\pi z_{2}z_{3}\alpha _{f}\sqrt{m_{0}c^{2}/\left(
2a_{123}\Delta \right) }$ with $a_{123}=A_{1}\left( A_{2}+A_{3}\right)
/\left( A_{1}+A_{2}+A_{3}\right) $. In the index $^{\prime }$reaction$%
^{\prime }$ the reaction resulting the two fragments will be marked (see
Table I.).

It is plausible to extend the investigation to the atomic gas-solid (e.g.
wall) interaction. In this case the role of particle $1$ is played by one
atom of the solid (metal) which is supposed to be formed from atoms with
nuclei of charge and mass numbers $z_{1}$and $A_{1}$. For initial state a
Bloch-function of the form
\begin{equation}
\varphi _{\mathbf{k}_{1,i}}(\mathbf{r}_{1})=N_{1}^{-1/2}\sum_{\mathbf{L}}e^{i%
\mathbf{k}_{1,i}\cdot \mathbf{L}}a(\mathbf{r}_{1}-\mathbf{L})
\end{equation}%
is taken, that is localized around all of the lattice points \cite{Ziman}.
Here $\mathbf{r}_{1}$ is the coordinate, $\mathbf{k}_{1,i}$ is wave number
vector of the first Brillouin zone ($BZ$) of the reciprocal lattice, $a(%
\mathbf{r}_{1}-\mathbf{L})$ is the Wannier-function, which is independent of
$\mathbf{k}_{1,i}$ within the $BZ$ and is well localized around lattice site
$\mathbf{L}$. $N_{1}$ is the number of lattice points of the lattice of
particles $1$. Repeating the cross section calculation applying
Bloch-function it is obtained that cross section results remain unchanged
and $n_{1}=N_{\text{1c}}/v_{\text{c}}$, where $v_{\text{c}}$ is the volume
of elementary cell of the solid and $N_{\text{1c}}$ is the number of
particles $1$ in the elementary cell.

\subsection{Numerical values of cross sections}

The cross section $\sigma _{23}^{\left( 2\right) }$ of the process $%
_{z_{1}}^{A_{1}}V+p+d\rightarrow $ $_{z_{1}}^{A_{1}}V^{\prime }+$ $%
_{2}^{3}He+5.493$ MeV is $\sigma _{23}^{\left( 2\right) }=n_{1}S_{\text{pd}%
}/v_{23}$, where $S_{\text{pd}}=1.89\times 10^{-53}z_{1}^{2}$ cm$^{6}$s$%
^{-1} $ with $z_{1}$ the charge number of the assisting nucleus. $\sigma
_{23}^{\left( 2\right) }$, similarly to thermal neutron capture cross
sections, has $1/v_{23}$ dependence. In case of $0.1$ eV initial kinetic
energy ($T=1160$ K if $kT=0.1$ eV) and with $z_{1}=54$ (Xe) $\sigma
_{23}^{\left( 2\right) }=n_{1}\times 2.5\times 10^{-31}$ b from which $%
\sigma _{23}^{\left( 2\right) }=0.0066$ nb at $n_{1}=2.65\times 10^{19}$ cm$%
^{-3}$ (which equals the number density of an atomic gas in normal state).
This value of $\sigma _{23}^{\left( 2\right) }$ is $10-15$ orders of
magnitude less than the thermal neutron capture cross sections.

In anomalous electron screening investigations accelerator of low energy
beams, e. g. in case of \cite{Huke} an accelerator line powered by a highly
stabilized 60-kV supply is applied. The targets are deuterium implanted
metals. Since our model is valid if the magnitude of initial kinetic
energies of particles $j=1-3$ are negligible compared to the reaction energy
$\Delta $, it can be applied. In this case in our model the role of particle
$1$ is played by one atom of the solid (metal). We focus on the $d(d,t)p$
reaction investigated in \cite{Huke} and we compare the cross section of the
assisted, second order process $\sigma _{23}^{\left( 2\right) }$ to the
cross section $\sigma _{23}^{\left( 1\right) }$ of the usual reaction. We
take $Pd$ as host metal. $v_{\text{c}}\left( Pd\right) =d^{3}/4\ $since $Pd$
has $fcc$ crystal structure and $N_{\text{1c}}=2$ resulting $n_{1}=N_{\text{%
1c}}/v_{\text{c}}=1.36\times 10^{23}$ cm$^{-3}$ ($d(Pd)=3.89\times 10^{-8}$
cm). We have calculated $S_{\text{d(d,t)p}}(Pd)$ taking $z_{1}=46$ and $%
A_{1}=106$ producing $S_{\text{d(d,t)p}}(Pd)=7.9\times 10^{-49}$ cm$^{6}$s$%
^{-1}$ and$\ n_{1}S_{\text{d(d,t)p}}\left( Pd\right) =1.08\times 10^{-25}$ cm%
$^{3}$s$^{-1}$. Taking $v_{23}=c\sqrt{2\varepsilon /\left( m_{0}c^{2}\right)
}$, $S\left( 0\right) =0.0571$ MeVb (see Table I.) and $2\pi \eta _{23}=2\pi
\alpha _{f}\sqrt{m_{0}c^{2}/\left( 2\varepsilon \right) }=$ $0.990/\sqrt{%
\varepsilon (\text{in MeV})}$ one obtains $\sigma _{23}^{\left( 2\right)
}=n_{1}S_{\text{d(d,t)p}}/v_{23}=7.77\times 10^{-11}/\sqrt{\varepsilon (%
\text{in MeV})}$ b\ and $\sigma _{23}^{\left( 1\right) }=0.0571\exp (-0.990/%
\sqrt{\varepsilon (\text{in MeV})})/\varepsilon ($in MeV$)$ b from $\left( %
\ref{sigma}\right) $. If $\sigma _{23}^{\left( 2\right) }>\sigma
_{23}^{\left( 1\right) }$ then the second order process dominates, i.e. if $%
7.35\times 10^{8}\exp (-0.990/\sqrt{\varepsilon })/\sqrt{\varepsilon }<1$
which is the case if $\varepsilon <0.001762$ MeV. Consequently the anomalous
screening phenomenon may be connected to the processes discussed here.
Moreover the experimental difficulties which accompanied anomalous screening
investigations indicate that the phenomenon discussed by us is difficult to
observe and examine, and partially answers the question why it was not
observed up till now.

\subsection{Experimental proposal}

The ground of the method which seems to be capable to show and to
investigate in detail the phenomenon may be the measurement of the assisting
particle and one from the two reaction products of e.g. metal assisted $%
d(d,t)p$ reaction in coincidence.

For this it is useful to determine the differential cross section
\begin{equation}
\frac{d\sigma _{23}^{\left( 2\right) }}{dEd\Omega }=F(E)=\frac{n_{1}}{v_{23}}%
\frac{A_{1}\alpha _{f}^{2}z_{1}^{2}}{\pi a_{23}a_{123}^{4}}\frac{S(0)c}{%
m_{0}c^{2}\Delta }\left( \frac{\hbar c}{\Delta }\right) ^{3}\chi \left[
x\left( E\right) \right] ,  \label{dsigma23-2}
\end{equation}%
where $x\left( E\right) =k_{1}/k_{m}=\sqrt{A_{1}E/\left( a_{123}\Delta
\right) }$ with $k_{m}=\sqrt{2m_{0}c^{2}a_{123}\Delta }/\left( \hbar
c\right) $, $k_{1}=\left\vert \mathbf{k}_{1}\right\vert $, $E$ and $\mathbf{k%
}_{1}$ are the energy and wave vector of particle $1$, and
\begin{equation}
\chi (x)=\left( \sum_{s=2,3}\frac{z_{s}a_{1s}\sqrt{A_{s}}}{\sqrt{%
e^{b_{23}A_{s}\frac{1}{x}}-1}}\right) ^{2}\frac{\sqrt{1-x^{2}}}{x^{8}}.
\label{dI/dx}
\end{equation}

\begin{figure}[tbp]
\resizebox{8.0cm}{!}{\includegraphics*{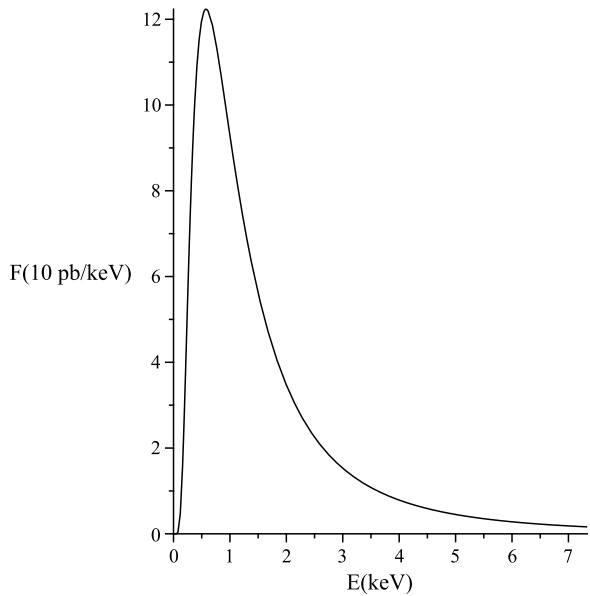}}
\caption{The energy $\left( E\right) $ dependence of the differential cross
section $d\protect\sigma _{23}^{\left( 2\right) }/\left( dEd\Omega \right)
=F\left( E\right) $ of $d(d,t)p$ reaction in the case of deuterized $Pd$. $F$
is given in $10$ pb/keV units. $E$ is the energy of the assisting particle $%
1 $ (in this case $Pd$) in keV units. The accelertor potential $U=1$ keV.
The possible maximum value of $E_{\text{.}}$is $146.65$ keV. }
\end{figure}

Fig. 1. shows the $E$ dependence of the differential cross section $d\sigma
_{23}^{\left( 2\right) }/\left( dEd\Omega \right) =F(E)$. If $\Phi _{2}$ is
the incident flux of particles $2$ then $dN_{1}/dt=N_{3}F(E)\Phi _{2}\delta
E\delta \Omega $ is the rate of particles $1$ of energy in the energy
interval $E\pm \delta E/2$ emitted in solid angle $\delta \Omega $ around
the direction determined by $\mathbf{k}_{1}$. $N_{3}$ is the total number of
particles $3$ irradiated by the beam of flux $\Phi _{2}$. It can be seen
from Fig. 1. that particles $1$ have kinetic energy mostly below $7$ keV.
Thus the wave vectors ($\mathbf{k}_{4}$ and $\mathbf{k}_{5}$) of the other
two final particles $4$ and $5$ have approximately opposite direction. Their
kinetic energies ($E_{4}$ and $E_{5}$) are peaked around $3\Delta /4$ and $%
\Delta /4$.

The accelerating electric potential $U$ seems to be worth decreasing below $%
1 $ keV since $d\sigma _{23}^{\left( 2\right) }/\left( dEd\Omega \right) $ $%
\sim 1/v_{23}\sim 1/\sqrt{U}$. Furthermore decreasing $U$ admits higher
accelerator current compared to the maximum of possible current of low
energy accelerators used in anomalous screening experiments \cite{Huke}.
However, decreasing $U$ results decreasing penetration depth of the beam
leading to decreasing interaction volume so that the optimal value of $U$
needs further study.

\section{Rate and power densities}

\begin{table}[tbp]
\tabskip=8pt
\centerline {\vbox{\halign{\strut $#$\hfil&\hfil$#$\hfil&\hfil$#$
\hfil&\hfil$#$\hfil&\hfil$#$\hfil&\hfil$#$\cr
\noalign{\hrule\vskip2pt\hrule\vskip2pt}
Reaction& S(0)&S_{reaction}&\Delta&p_{reaction}\cr
\noalign{\vskip2pt\hrule\vskip2pt}
d(d,n)_{2}^{3}He & 0.055 & 1.01\times10^{-48} & 3.269 &9.82 \cr
d(d,p)t& 0.0571 & 1.10\times 10^{-48}&4.033& 13.2\cr
d(t,n)_{2}^{4}He & 11.7 & 1.06\times 10^{-46} &17.59& 5.57\times 10^{3} \cr
_{2}^{3}He(d,p)_{2}^{4}He & 5.9 & 1.51\times 10^{-48}&18.25& 82.6\cr
_{3}^{6}Li(p,\alpha)_{2}^{3}He & 2.97 & 1.99\times10^{-49} & 4.019 & 2.38 \cr
_{3}^{6}Li(d,\alpha)_{2}^{4}He & 16.9 & 1.33\times10^{-49} & 22.372 & 8.84 \cr
_{3}^{7}Li(p,\alpha )_{2}^{4}He & 0.0594 & 3.85\times10^{-51} & 17.347 & 0.199 \cr
_{4}^{9}Be(p,\alpha)_{3}^{6}Li & 17 & 1.79\times10^{-49} & 2.126 & 1.13 \cr
_{4}^{9}Be(p,d)_{4}^{8}Be & 17 & 1.66\times10^{-49} & 0.56 &0.277 \cr
_{4}^{9}Be(\alpha ,n)_{6}^{12}C & 2.5\times10^{3} & 6.22\times10^{-51} & 5.701 & 0.106 \cr
  & 6\times10^{5} & 1.49\times10^{-48} & 5.701 & 25.4 \cr
_{5}^{10}B(p,\alpha )_{4}^{7}Be & 4 & 1.04 \times10^{-50} & 1.145 & 0.0356 \cr
  & 2\times10^{3} & 5.21\times10^{-48} & 1.145 & 17.8 \cr
_{5}^{11}B(p,\alpha )_{4}^{8}Be & 187 & 5.16\times10^{-49} & 8.59 & 13.2 \cr
\noalign{\vskip2pt\hrule\vskip2pt\hrule}}}}
\caption{$S_{\text{reaction}}$ and power density of $Xe$ assisted reactions
with two final fragments in long wavelength approximation. $S(0)$ is the
astrophysical $S$-factor at $\protect\varepsilon =0$ in MeVb \protect\cite%
{Angulo}, \protect\cite{Descou}, \protect\cite{Spitaleri}. $S_{\text{reaction%
}}$ (in cm$^{6}$s$^{-1}$) is calculated using $\left( \protect\ref{result2}%
\right) $ with $\left( \protect\ref{Iintcharged}\right) $ taking $z_{1}=54$ $%
\left( Xe\right) $, $\Delta $ is the energy of the reaction in MeV and $p_{%
\text{reaction}}=n_{1}n_{2}n_{3}S_{\text{reaction}}\Delta $ is the power
density in Wcm$^{-3}$ that is calculated with $n_{1}=n_{2}=n_{3}=2.65\times
10^{20}$ cm$^{-3}$. In the case of $_{4}^{9}Be(\protect\alpha ,n)_{6}^{12}C$
and $_{5}^{10}B(p,\protect\alpha )_{4}^{7}Be$ reactions the astrophysical $S$%
-factor $[S(\protect\varepsilon )]$ has strong energy dependence therefore
the calculation was carried out with two characteristic values of $S(\protect%
\varepsilon )$. }
\end{table}

The rate in volume $V$ is
\begin{equation}
\frac{dN_{\text{reaction}}}{dt}=N_{3}\Phi _{23}\sigma _{23}^{\left( 2\right)
},
\end{equation}%
where $\Phi _{23}=n_{2}v_{23}$ is the flux of particles $2$ with $n_{2}=$ $%
N_{2}/V$ their number density. $N_{2}$ and $N_{3}$ are the numbers of
particles $2$ and $3$ in the normalization volume. The rate and power
densities are defined as
\begin{equation}
r_{\text{reaction}}=\frac{1}{V}\frac{dN_{\text{reaction}}}{dt}%
=n_{3}n_{2}n_{1}S_{\text{reaction}}
\end{equation}%
and%
\begin{equation}
p_{\text{reaction}}=r_{\text{reaction}}\Delta =n_{1}n_{2}n_{3}S_{\text{%
reaction}}\Delta ,
\end{equation}%
respectively, where $n_{3}=N_{3}/V$ is the number density of particles $3$. $%
r_{\text{reaction}}$ and $p_{\text{reaction}}$ are both temperature
independent.

The rate $\left( r_{\text{pd}}\right) $ and power $\left( p_{\text{pd}%
}\right) $ densities of reaction $_{z_{1}}^{A_{1}}V+p+d\rightarrow $ $%
_{z_{1}}^{A_{1}}V^{\prime }+$ $_{2}^{3}He$ are determined taking $z_{1}=54$ (%
$Xe$) and $n_{1}n_{2}n_{3}=1.86\times 10^{61}$ cm$^{-9}$(which is the case
e.g. at $n_{1}=n_{2}=n_{3}=2.65\times 10^{20}$ cm$^{-3}$, $n_{1}$, $n_{2}$
and $n_{3}$ are the number densities of $Xe$, $p$ and $d$, i.e. particles 1,
2 and 3) for which considerable values are obtained: $r_{\text{pd}%
}=1.02\times 10^{12}$ cm$^{-3}$s$^{-1}$ and $p_{\text{pd}}=0.901$ Wcm$^{-3}$%
. If the impurity is $Hg$ or $U$ then these numbers must be multiplied by $%
2.2$ or $2.9$, respectively.

The results of $S_{\text{reaction}}$ and power density calculations of a
number of $Xe$ assisted reactions with two final fragments in long
wavelength approximation and with $n_{1}=n_{2}=n_{3}=2.65\times 10^{20}$ cm$%
^{-3}$ can be found in Table I.

To reach the order of magnitude $10^{61}$ cm$^{-9}$ of $n_{1}n_{2}n_{3}$ is
a great challenge. It may be done e.g. with the aid of dissociative
chemisorption at metal (e.g. $Pd$, $Ni$ and $Cu$) surfaces from two atomic
molecules, e.g. $H_{2}$, $HD$ or $D_{2}$ by heating molecular gas \cite%
{Kroes}. In this case $n_{1}>10^{22}$ cm$^{-3}$ is the number density of
metal atoms in the solid and $n_{1}n_{2}n_{3}=10^{61}$ cm$^{-9}$ can be
reached if $kT\sim 0.5-1$ eV producing $n_{2}=n_{3}>5.3\times 10^{19}$ cm$%
^{-3}$. It can be achieved in a two atomic gas in the $4-8$ atm pressure, $%
600-1200$ K temperature range, respectively, at the surface. In the case of
powdered samples of small grain size or nanoparticles one may reach
interaction volume large enough to be able to generate heat produced by
power densities of some of nuclear reactions listed in Table I. that\ is
observable with the aid of precise calorimetric measurements.

Since in $\left( \ref{Reaction 3}\right) $ and $\left( \ref{Reaction 4}%
\right) $ the reaction energy is taken away by particles $%
_{z_{1}}^{A_{1}}V^{\prime }$, $_{z_{3}+z_{2}}^{A_{3}+A_{2}}Y$ and $%
_{z_{1}}^{A_{1}}V^{\prime }$, $_{z_{4}}^{A_{4}}Y$, $_{z_{5}}^{A_{5}}W$,
respectively, as their kinetic energy that they lose in a very short range
to their environment converting the reaction energy efficiently into heat if
the state of matter of atomic state is dense, so their direct observation is
difficult in this case.

In the experimental conditions stated above the creation of new elements due
to nuclear reactions i.e. the presence of nuclear transmutation in the
system may be a way to confirm our predictions experimentally.

\section{Conclusion}

It is found that \textit{any perturbation }may lead to nonzero cross section
and rate of nuclear reactions forbidden in the $\varepsilon \rightarrow 0$
limit. Since this statement applies to every nuclear process forbidden in
the $\varepsilon \rightarrow 0$ limit it concerns low energy nuclear physics
with charged participants in general. Thus, it may be stated that a very
great number of reactions, which are determined by different initial states,
different perturbations and different processes of second and higher order
and which may be attached to forbidden reactions, have not been investigated
up till now.


\begin{thebibliography}{99}
\bibitem{Angulo} C. Angulo, M. Arnould, M. Rayet, P. Descouvemont, D. Baye,
C. Leclercq-Willain, A. Coc, S. Barhoumi, P. Aguer, C. Rolfs, R. Kunz, J. W.
Hammer, A. Mayer, T. Paradellis, S. Kossionides, C. Chronidou, K. Spyrou, S.
Degl'Innocenti, G. Fiorentini, B. Ricci, S. Zavatarelli, C. Providencia, H.
Wolters, J. Soares, C. Grama, J. Rahighi, A. Shotter, and M. Lamehi Rachti,
Nucl.Phys. A \textbf{656}, 3-183 (1999).

\bibitem{Huke} A. Huke, K. Czerski, P. Heide, G. Ruprecht, N. Targosz, and
W. Zebrowski, Phys. Rev. C \textbf{78}, 015803 (2008).

\bibitem{FP1} M. Fleishmann and S. Pons, J. Electroanal. Chem. \textbf{261},
301-308 (1989).

\bibitem{Huizenga} J. R. Huizenga, Cold Fusion: The Scientific Fiasco of the
Century (University of Rochester Press, Rochester, 1992).

\bibitem{Krivit} S. B. Krivit and J. Marwan, J. Environ. Monit. \textbf{11}
1731-46 (2009).

\bibitem{Storms3} E. Storms, \textit{The Science of Low Energy Nuclear
Reaction, A Comprehensive Compilation of Evidence and Explanations about
Cold Fusion} (World Scientific, Singapore, 2007).

\bibitem{Storms2} E. Storms, Naturwissenschaften \textbf{97}, 861-881 (2010).

\bibitem{Storms1} E. Storms, Current Science\textit{\ }\textbf{108,} 535
(2015).

\bibitem{Alder} K. Alder, A. Bohr, T. Huus, B. Mottelson, and A. Winther,
Rev. Mod. Phys. \textbf{28}, 432-542 (1956).

\bibitem{Blatt} J. M. Blatt, and V. F. Weisskopf, \textit{Theoretical
Nuclear Physics} (Wiley, New York, 1952).

\bibitem{Bethe} H. A. Bethe and E. E. Salpeter, \textit{Quantum Mechanics of
One- and Two Electron Atoms} (Springer, Heidelberg, 1957), sec. 67.

\bibitem{Shir} R. B. Firestone and V. S. Shirly, \textit{Tables of Isotopes}%
, 8th ed. (Wiley, New York, 1996).

\bibitem{Kroes} G. J. Kroes, A. Gross, E. J. Baerends, M. Schefler, and D.
A. McCormack, Acc. Chem. Res. \textbf{35},\ 193-200 (2002).

\bibitem{Landau} L. D. Landau and E. M. Lifsic, \textit{Course of
Theoretical Physics,} Vol. 3., \textit{Quantum Mechanics: Non-relativistic
Theory,} 3rd, revised Ed., (Pergamon, Oxford, 1991) p. 154-156 (Translated
from 4th Ed. of 'Kvantovaya mechanika: nerelyativistskaya teorija,
Izdatel'stvo "Nauka" Moskow, 1989').

\bibitem{Ziman} J. M. Ziman, \textit{Principles of the Theory of Solids}
(University Press, Cambridge, 1964).

\bibitem{Descou} P. Descouvemont, A. Adahchour, C. Angulo, A. Coc, and E.
Vangioni-Flam, At. Data Nucl. Data Tables, \textbf{88}, 203-236 (2004).

\bibitem{Spitaleri} C. Spitaleri, S. Typel, R. G. Pizzone, M. Aliotta, S.
Blagus, M. Bogovac, S. Cherubini, P. Figuera, M. Lattuada, M. Milin, D.
Miljani\'{c}, A. Musumarra, M. G. Pellegriti, D. Rendi\'{c}, C. Rolfs, S.
Romano, N. Soi\'{c}, A. Tumino, H. H. Wolter, and M. Zadro, Phys. Rev. C
\textbf{63}, 055801 (2001).
\end{thebibliography}
\end{document}